# Assessing Suburban Air Quality Constraints on Free Cooling in an Irish City

Paul D. O' Sullivan, PhDTheofanis Psomas, PhDGanga C. R. Devarapu, PhD

Adam C. O' Donovan, PhD

**ABSTRACT**

*Temperate climates are expected to have an expansion of the number of hours where ventilative cooling is needed and where potential is available even under climate change scenarios. Further to this, the use of free cooling in the form of untreated outdoor air has the potential to be a renewable energy source that adds flexibility to national electricity grids if used appropriately and is critical in decarbonising cooling more generally, particularly where coupled with significant thermal mass in buildings. The following study utilises external air quality measurements from the PASSESPARTOUT project and were collected from unique external air quality systems located in Cork in Ireland. The study characterises this data according to constraints based on control limits for cooling systems and for individual external air quality parameters. While the work aims to demonstrate the barriers to cooling potential that may exist due to poor external air quality, it also offers insights into where good urban air quality supports diurnal ventilative cooling strategies that are best suited given these limitations. The findings show that out of seven outdoor pollutants investigated, $NO_2$ and $PM_{2.5}$ were consistently above recommended threshold limits in suburban areas and present a barrier to the use of indoor-outdoor direct airflow coupling using natural ventilation strsategies for ventilative cooling purposes. Indoor concentration levels within the surrounding areas to those investigated should be evaluated to develop Indoor-Outdoor ratios for designers for barrier pollutants.*

## INTRODUCTION

Ventilating indoor spaces using untreated outdoor ambient air has many benefits (Najafi Ziarani et al., 2023; O' Donovan et al., 2021; O' Donovan and O' Sullivan, 2023; Sohail et al., 2024). Its supply has a low carbon footprint, and it can, when conditions are right, act as a contaminant removal mechanism. One of the main attractions of using ambient outdoor air to directly ventilate indoor spaces is its heat removal potential, or its role as resilient ventilative cooling (VC) in maintaining acceptable indoor thermal environments in a low carbon way (Tavakoli et al., 2022). In fact, the additional airflow above the minimum requirements for hygienic/pollutant removal purposes that is used for cooling, utilising the thermal energy store in the ambient air, is recognized as a renewable source in the EU Renewable Energy Directive (European Commission, 2021, 2018). The drawback with using VC, i.e. the cooling potential in untreated outdoor air that has not been conditioned using mechanical cooling, is the barriers to its use which restricts access to cooling potential at times when other unacceptable conditions are present. These include excessive ambient noise, high levels of outdoor air pollution, risk of unwarranted access to buildings through openings, high winds and rain, amongst others (Carrilho Da Graça and Linden, 2016). The focus of the research presented in this paper was to evaluate outdoor air quality as a barrier to the use of the free cooling potential available in the ambient air. The objective of the research was to assess whether outdoor air pollutant levels were too high in

**Paul D. O' Sullivan** is a lecturer and principal investigator at the MeSSO Research Group, in the Department of Process, Energy and Transport Engineering, at Munster Technological University, Ireland. **Theofanis Psomas** is a researcher at the MeSSO Research Group, in the Department of Process, Energy and Transport Engineering, at Munster Technological University, Ireland. **Ganga Chinna Rao Devarapu** is a research fellow at the Centre for Advanced Photonics & Process Analysis (CAPPA) at Munster Technological University, Ireland. **Adam O' Donovan** is a senior researcher at the MeSSO Research Group, in the Department of Process, Energy and Transport Engineering, at Munster Technological University, Ireland.

a suburban mild climate, well suited to natural VC, to allow direct coupling of the indoor and outdoor air using natural ventilation strategies for cooling purposes, thus requiring mechanical ventilation with in-line air filtration or sub zone portable air cleaners. For pollutants $NO_2$ and PM2.5, while the WHO and EU have guidelines for thresholds, there seems to be no agreed threshold below which there is no harm to occupants. For this reason, we are particularly interested in indoor exposure to these from coupling the indoor and outdoor air directly using natural ventilative cooling. The PASSEPARTOUT research project is aiming to develop compact, photonic-based gas analysers for a smart sensing solution to environmental pollution monitoring in urban areas (PASSEPARTOUT Consortium, 2025). It combines expertise in lasers, spectroscopy, data analysis, systems integration, environmental testing and drone operations in a Europe-wide consortium involving both academia and industry.

The data used in this study has been gathered within the PASSEPARTOUT project. The materials and methods section summarises details about the data gathering and thresholds adopted for assessment of outdoor air quality as a barrier to VC while results and discussion section evaluates the data and discusses any substantive insights.

## MATERIALS AND METHODS

External air quality data was collected from eight different locations in suburban areas around Cork City, Ireland. These locations included: Bishopstown, Carrigtwohill, Douglas, Mayfield, Knockrea, Midleton, Carrigtwohill and Little Island. The air quality monitoring system used (known as AirSENCE, see Figure 1) measured Nitric Oxide (NO), Nitrogen Dioxide ($NO_2$), Carbon Monoxide (CO), Ozone ($O_3$), Sulphur Dioxide ($SO_2$), Particulate Matter; < 10 Micrometers ($PM_{10}$), < 2.5 Micrometers ($PM_{2.5}$), < 1 Micrometer ($PM_1$), Air Temperature and Relative Humidity, at a 1-minute sampling rate. Data was collected from June 2022 to May 2024. The accuracy of the AirSENCE units is indicated in Table A.1 (see Appendix). Table 1 indicates the thresholds used for hourly, daily and annual analyses utilizing limits according to World Health Organization (WHO) guidelines which were updated in 2021 (WHO, 2021) as well practical limits on the use of natural ventilation (maximum external temperature).

Table 1. Table of thresholds used to assess data

| Parameter (units) | Averaging Time | Limits |
|---|---|---|
| $PM_{2.5}$ (µg/m$^3$) | Annual | 5 |
|  | 24-hour | 15 |
|  | Hourly | 15 |
| $PM_{10}$ (µg/m$^3$) | Annual | 15 |
|  | 24-hour | 45 |
|  | Hourly | 45 |
| $O_3$ (µg/m$^3$) | Peak season | 60 |
|  | 8-hour | 100 |
|  | Hourly | 100 |
| $NO_2$ (µg/m$^3$) | Annual | 10 |
|  | 24-hour | 25 |
|  | Hourly | 25 |
| $SO_2$ (µg/m$^3$) | 24-hour | 40 |
|  | Hourly | 40 |
| CO (mg/m$^3$) | 24-hour | 4 |

Finally, limits from EN16798-1 were used for acoustic barriers (NSAI, 2019). Table 1 indicates the thresholds used to characterize and contextualise this data. The work presented here is similar to the work of Belias et al. (Belias and Licina, 2023), however it considers temperature in a similar manner to Bravo-Diaz et al. (Bravo Dias et al., 2020). Data was converted from parts per billion to the units shown in Table 1 for $NO_2$, CO, $O_3$ and $SO_2$, using the conversion factors by the Irish Environmental Protection Agency (EPA, 2025). In this paper, two separate analyses were conducted; 1) assessment of the outdoor air quality (OAQ) with respect to annual (using all available data) and 24-hour limits indicated in Table 1, and 2) assessment of the effect that outdoor air quality will have on the potential for ventilative cooling (VC) in Cork. For this stage, 24-hour OAQ limits were adopted as the threshold to assess whether external conditions were suitable for ventilative cooling. Data was analysed at an hourly level during external conditions conductive for VC (by accessing whether VC was possible in Cork using Cork Airport data from Met Éireann data (Met Eireann, 2025)).

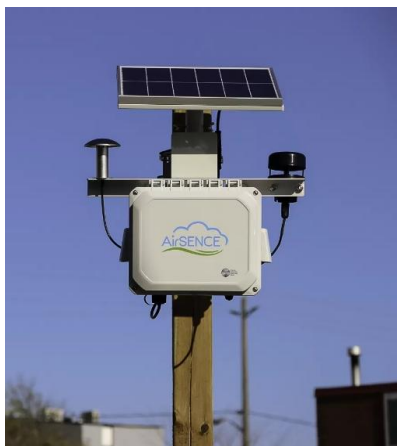

**Figure 1** Image of AirSENCE (AirSENCE, 2025) external air quality sensors deployed in all locations.

## RESULTS AND DISCUSSION

### External air quality – 24-hour and overall observations

Figure 1 and Table 2 highlight the overall and daily average OAQ conditions in each location and for each relevant parameter where limits by the WHO exist and apply. Annually, all locations do not comply with the WHO guidelines for $PM_{2.5}$, however, all but one (L170) comply with $PM_{10}$ limits. All locations had annual averages that were less than thresholds for $SO_2$ and $O_3$. However, all locations had annual averages that exceeded thresholds for $NO_2$. This result was is consistent with values also indicated in the EPA's annual report for 2023, which highlighted that Ireland met its EU requirements but not WHO requirements (EPA, 2024). It should be noted that the changes in air quality guidelines (AQGs) between 2005 and 2021 have led to a categorical change in the conclusions that can be drawn on the effect that $PM_{2.5}$ and $NO_2$ have on the use of natural ventilation.

**Table 2 – Average outdoor air quality in suburban Cork for sampling period, N. (\* indicates non-compliance with WHO thresholds)**

| Location | $PM_1$ | $PM_{2.5}$ | $PM_{10}$ | CO | $O_3$ | $NO_2$ | $SO_2$ | N (months) |
|---|---|---|---|---|---|---|---|---|
| L164 | 5.6 | 8.0* | 9.6 | 0.3 | 54.2 | 18.7* | 5.0 | 11 |
| L165 | 5.0 | 6.9* | 8.9 | 0.3 | 44.6 | 18.9* | 4.7 | 8 |
| L166 | 4.6 | 6.0* | 7.4 | 0.3 | 44.3 | 19.1* | 9.5 | 6 |
| L167 | 3.8 | 5.2* | 5.9 | 0.4 | 22.9 | 29.6* | 13.9* | 6 |
| L168 | 4.6 | 6.5* | 7.6 | 0.3 | 56.7 | 36.4* | 4.1 | 6 |
| L169 | 5.5 | 7.3* | 8.6 | 0.4 | 75.3 | 35.8* | 5.9 | 7 |
| L170 | 27.2 | 30.3* | 33.4* | 0.2 | 34.7 | 41.9* | 6.6 | 4 |
| L171 | 4.7 | 6.4* | 7.8 | 0.3 | 48.0 | 22.0* | 4.3 | 11 |

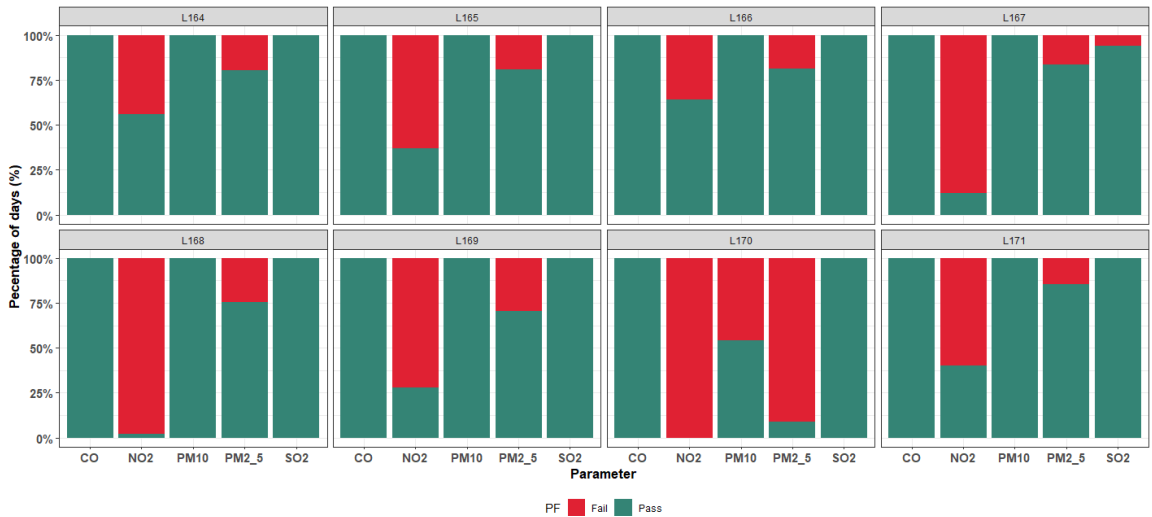

**Figure 2** Percentage of days where average measurements of each parameter exceed the daily thresholds shown in Table 1.

On a daily basis, between 1 to 153 days exceeded OAQ thresholds that were parameter and location dependent. Only one location failed daily $SO_2$ limits (L167). All locations except one had $PM_{10}$ levels that were deemed acceptable (L170), while between 5% and 74% of days failed PM2.5 levels. Finally, between 26% and 100% of days in each location exceeded daily $NO_2$ thresholds. There appears to be a challenge for ventilation designers to utlise natural ventilation in suburban areas in Cork City while also limiting the growth of PM2.5 and $NO_2$ indoors. There are multiple locations with $PM_{2.5}$ exceeding daily threshold values for 20% of total monitored days. This is a significant proportion of these days in the extended cooling season now evident with low energy, high performance Nearly Zero Energy Buildings in mild climates such as Irelands, i.e. from April to October.

### Barriers to natural ventilation – hourly observations

The potential to cool in Cork in 2023 and 2024 was quite high, there were only four days in Cork airport where the daily maximum air temperature exceeded 23°C.

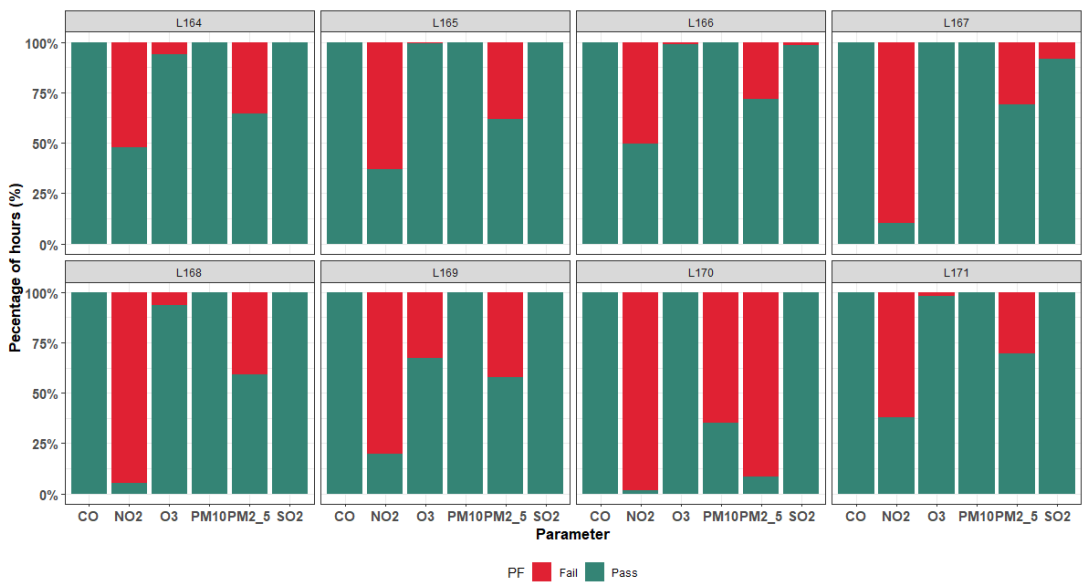

**Figure 3** Percentage of hours where average measurements of each parameter exceed thresholds shown in Table 1.

The following results are presented on the assumption that by and large the entire year is acceptable for natural ventilation. Figure 3 indicates the percentage of total hours where conditions were considered unacceptable according to WHO guidelines in line with the work of (Belias and Licina, 2023), however, considering new updated limits from 2021. Taking each parameter initially, CO levels were seen as being acceptable in all locations. $SO_2$ levels reduced the number of cooling hours by 0% to 2%.

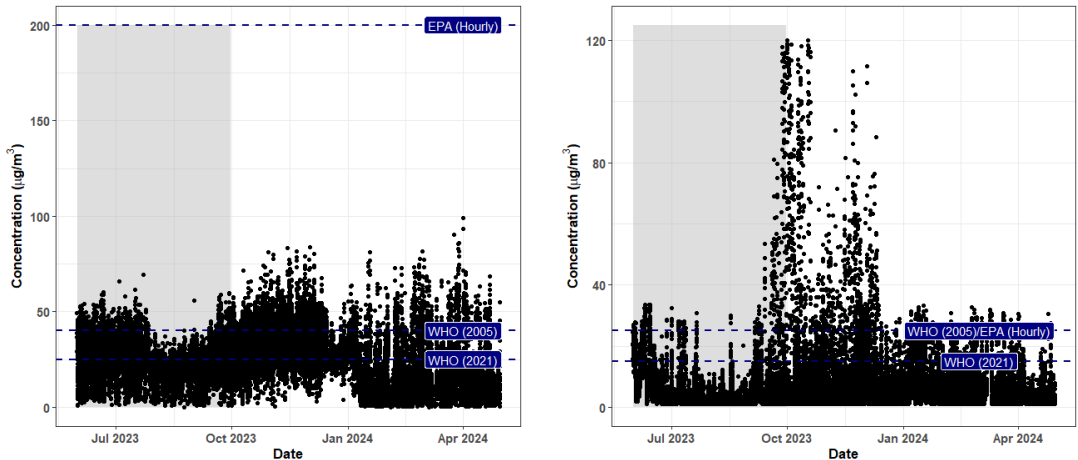

**Figure 4** Scatterplot of concentration of different outdoor polluants in all measurement locations (Left: indicates concentration of $NO_2$, Right: indicates concentration of $PM_{2.5}$, shaded area indicates the typical cooling season, dashed lines highlight different thresholds from different sources)

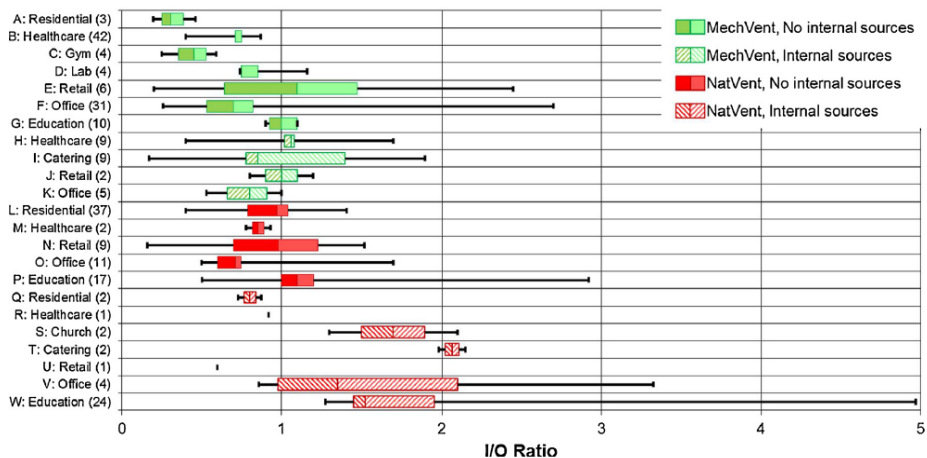

**Figure 5** Reproduced from (Martins and Carrilho Da Graça, 2018): Selection of studies with PM2.5 I/O ratios by building typology (number in parenthesis indicates sample size), ventilation type and existence of internal sources; the box plot represents minimum, 25th percentile, median, 75th percentile and maximum.

$O_3$ reduced the number of cooling hours by 0% to 24% depending on the location. $NO_2$ levels reduced the number of cooling hours by 32% to 96% depending on the location. $PM_{10}$ was seen as acceptable in all locations except for one location where $PM_{10}$ reduced the number of cooling hours by 29%. $PM_{2.5}$ reduced the number of available cooling hours in all locations, ranging from 8% to 59% depending on location. It should be noted that different threshold limits have been applied by different agencies and researchers in literature. Figure 4 highlights the variation in the concentration of different pollutants over time for all locations. While the EPA reports an upper limit over 18 hours that is quite high (200μg/m$^3$) and would increase the number of hours where natural ventilation could be used, the concentrations of $NO_2$ that have been measured in all sites show a need to mitigate the source of $NO_2$. The concentrations of both $NO_2$ and $PM_{2.5}$ are less in the typical cooling season, however, the likely trajectory in terms of OAQ thresholds presents a significant barrier in the use of natural ventilation systems A recent review of the impact of $PM_{2.5}$ in indoor urban environments presented indoor-Outdoor Ratios for $PM_{2.5}$ for different ventilation strategies adopted. Figure 5 is reproduced directly from this study. It is evident that natural ventilative strategies where no $PM_{2.5}$ sources exist indoors are susceptible to an I/O Ratio of around 1.0 suggesting a complete tranfer of $PM_{2.5}$ levels to indoors from the outdoor pollutant source, resulting in chronic long-term exposure for building occupants leading to acute health conditions. For this reason, pollutants should be considered by designers and building operators as barriers to ventilative cooling. This can present a significant challenge as regulating barriers such as those from OAQ are often beyond the control of building designers.

## CONCLUSIONS AND FUTURE WORK

The positive finding from the analysis of this data is that out of the seven outdoor pollutants investigated, only two had levels that could be considered as chronically above recommended threshold limits. However, reducing these to acceptable levels requires significant effort by authorities responsible for OAQ in suburban areas surrounding Cork, Ireland. Indoor-Outdoor coupling through direct path ventilation strategies is an important low carbon means to provide healthy and safe indoor spaces for occupants during times when there is a need to reduce indoor air temperatures. Future work should consider the effect that $NO_2$ and $PM_{2.5}$ has on the ability of buildings to ventilate naturally and evaluate interventions that can help reduce the outdoor concentration levels. Efforts should also evaluate indoor levels of the same pollutants at locations near to the AirSENCE monitoring points. This situation is likely further exacerbated in other Irish Cities, in particular, Dublin. Some work is needed to adequately quantify the extent of ambient thermal reserve exists for ventilative cooling and the nature and extent of outdoor air pollutant 'contamination' of this cooling thermal reserve.

## ACKNOWLEDGMENTS

The work was supported under the RESILIENCE project grant, supported by the Sustainable Energy Authority of Ireland under the Research, Development and Demonstration Grant Agreement 19/RDD/496.

## APPENDIX

**Table A1 – Accuracy of sensors used in AirSENCE units**

| Variable | Range | Detection Limit | Accuracy | Precision | Sampling rate |
|---|---|---|---|---|---|
| $PM_1$ | 0-6000 μg/m$^3$ | 5 μg/m$^3$ | <10 μg/m$^3$ @ <100 μg/m$^3$; <10% >100 μg/m$^3$ | ±10% | 5s |
| $PM_{2.5}$ | 0-6000 μg/m$^3$ | 5 μg/m$^3$ | <10 μg/m$^3$ @ <100 μg/m$^3$; <10% >100 μg/m$^3$ | ±10% | 5s |
| $PM_{10}$ | 0-6000 μg/m$^3$ | 5 μg/m$^3$ | <10 μg/m$^3$ @ <100 μg/m$^3$; <10% >100 μg/m$^3$ | ±10% | 5s |
| CO | 0-6000 ppb | 20 ppb | ±25ppb or ±5%, whichever is greater | - | 1s |
| $O_3$ | 0-10000 ppb | 5 ppb | ±5ppb or ±5%, whichever is greater | - | 1s |
| $NO_2$ | 0-10000 ppb | 5 ppb | ±5ppb or ±5%, whichever is greater | - | 1s |
| $SO_2$ | 0-10000 ppb | 5 ppb | ±10ppb or ±5%, whichever is greater | - | 1s |